\documentstyle[12pt]{article}
\newcommand{\be}{\begin{equation}}
\newcommand{\ee}{\end{equation}}
\newcommand{\ba}{\begin{eqnarray}}
\newcommand{\ea}{\end{eqnarray}}

\newcommand{\LL}{{\cal{L}}}

\date{}

\newcommand{\grgl}{\:\hbox to -0.2pt{\lower2.5pt\hbox{$\sim$}\hss}
           {\raise3pt\hbox{$>$}}\:}
\newcommand{\klgl}{\:\hbox to -0.2pt{\lower2.5pt\hbox{$\sim$}\hss}
           {\raise3pt\hbox{$<$}}\:}

\begin{document}
\begin{titlepage}
\begin{flushright}
HD--THEP--97--28 \\
SOGANG--HEP 222/97
\end{flushright}
\bigskip
\begin{center}
{\bf\LARGE Hamiltonian Embedding of SU(2) Higgs Model }\\
\vspace{0.5cm}
{\bf \LARGE in the Unitary Gauge}\\
\vspace{1cm}
Yong-Wan Kim\footnote{e-mail:~kim@thphys.uni-heidelberg.de
and yjpark@ccs.sogang.ac.kr
                                                                                
On leave of absence from 
Department of Physics and Basic Science Research Institute,
Sogang University, C.P.O.Box 1142,
Seoul 100-611, Korea}, 
Young-Jai Park$^1$
and Klaus D. Rothe\footnote{e-mail:~k.rothe@thphys.uni-heidelberg.de} \\
\bigskip
Institut  f\"ur Theoretische Physik\\
Universit\"at Heidelberg\\
Philosophenweg 16, D-69120 Heidelberg\\
\vspace{1cm}
\end{center}
\begin{abstract}
\noindent Following systematically the generalized Hamiltonian approach of  
Batalin, Fradkin and Tyutin (BFT), 
we embed the second-class non-abelian SU(2) 
Higgs model in the unitary gauge into a gauge invariant theory. 
The strongly involutive 
Hamiltonian and constraints are obtained as an infinite power series in the 
auxiliary fields. 
Furthermore, comparing these results with those obtained from the gauged 
second class Lagrangian, we arrive at a simple 
interpretation for the first class Hamiltonian, constraints and observables.
\end{abstract}
\vspace{2.5cm}
\end{titlepage}

\section{Introduction}
The quantization of second-class Hamiltonian systems \cite{Di}
 requires the strong implementation of the second-class constraints. This may  
imply Dirac brackets, whose non-canonical structure may pose problems on  
operator level. 
This makes it desirable to embed  the second-class theory into  
a first-class one, where the commutator relations remain canonical and the  
constraints are imposed on the states. 
An example is provided by the Higgs model with spontaneous
symmetry breakdown \cite{AL} whose quantization
is usually carried out in the so called ``unitary'' gauge. As is well known,
in this gauge the model is a purely second class system characterized by
two sets of second class constraints \cite{GR,BRR1}. The required strong
implementation of these constraints leads to non-polynomial field dependent
Dirac brackets. As mentioned above, 
one can circumvent the problems associated with this
non-polynomial dependence by turning this system into a first class one with 
a usual Poisson bracket structure in an extended phase space
and implementing the first class constraints on the physical states. 
A systematic procedure for achieving this has been given by Batalin and
Fradkin (BF) \cite{BF} and has been explicitly carried out for the above 
model for the abelian case \cite{BRR1}. 
The construction in the BF framework
proved already non-trivial in the abelian case, and does not appear          
particularly suited for treating the non-abelian case. 

In this paper we shall generalize the above program to the case of the 
non-abelian SU(2) Higgs model as a nontrivial and simple example
by following a simpler constructive procedure
based on modified version of the BF-formalism, due to Batalin and Tyutin
\cite{BT}. This procedure has been recently applied to several interesting
abelian \cite{BR,AD,KK} and non-abelian 
models \cite{BBN,PP,KR}, and will also render the analysis 
of the SU(2) Higgs model in the unitary gauge more transparent.

Starting from the second class Lagrangian, we systematically construct in
section 2 the corresponding first class, strongly involutive constraints, 
following the BFT-procedure. In section 3 we then use the same procedure
in order to construct the first-class fields in strong involution with
the constraints. The first class Hamiltonian is obtained from the
original second-class Hamiltonian by replacing the original fields by the
corresponding first class ones. This Hamiltonian is thus again in strong
involution with the constraints. In section 4 we then show that the
results coincide with those obtained by gauging the original second class
Lagrangian, after a suitable canonical transformation. This establishes
the equivalence of the BFT construction and the Lagrangian quantization
procedure based on the addition of a Wess-Zumino (WZ) term 
\cite{BSV,HT,AAR}
(see also \cite{BBN,PP,KR}). We conclude in section 5 with a summary.

\section{BFT construction of first class constraints}

Consider the non-Abelian SU(2) Higgs model in the unitary gauge,
\be\label{2.1}
\LL(B^\mu, \eta)= {1 \over 4} tr G_{\mu\nu}G^{\mu\nu} 
     - {1\over 2} m^2(\eta) tr B_\mu B^\mu
     +{1 \over 2}\partial_\mu \eta \partial^\mu \eta + V(\eta),
\ee
where $V(\eta)$ is the Higgs potential 
\be\label{2.2}
V(\eta) = {\mu^2 \over 2} (\eta + v)^2 - {\lambda \over 4} (\eta + v)^4
\ee
with the vacuum expectation value $v$,
and for later convenience we define
the field dependent mass squared $m^2(\eta)$ as follows
\be\label{2.3}
m^2(\eta) = (\eta + v)^2.
\ee
Here the vector potential $B^\mu$ is an antihermitian 
Lie algebra valued field
\be\label{2.4}
B^\mu =  t^a B^{\mu a} ~~~(a=1,2,3),
\ee
where $t^a$ are the SU(2) group generators,
and $G^{\mu\nu}$ is the associated field strength tensor
\be\label{2.5}
G^{\mu\nu}=\partial^\mu B^{\nu}-\partial^\nu B^{\mu} + [B^{\mu}, B^{\nu}]. 
\ee
Our conventions are
\ba\label{2.6}
[t^a,t^b]= \epsilon^{abc} t^c,~~~
tr(t^a t^b)=-\delta^{ab},\ea
where $\epsilon^{abc}$ is the structure constant of the SU(2) group.
The momenta canonically conjugate to $B^{0a}$, $B^{ia}$ and $\eta$ are 
respectively given by $\pi^a_0=0$, $\pi^a_i=G^a_{i0}$ and $\pi=\dot \eta$. 
We thus have the primary constraints $\pi^a_0=0$.
The canonical Hamiltonian density associated with the Lagrangian 
(\ref{2.1}) is found to be
\be\label{2.7}
{\cal H}_C= {1\over 2} (\pi^a_i)^2 
            + {1\over 2} \pi^2
            + {1 \over 4} (G^a_{ij})^2             
            + {1\over 2} m^2(\eta) (B^{ia})^2  
            + {1\over 2} m^2(\eta) (B^{0a})^2              
            + {1\over 2} (\partial_i \eta)^2
            - V(\eta) 
            - B^{0a} \Omega_2^a.
           \ee
Since persistency in time of these constraints 
leads to further (secondary) constraints $\Omega^a_2=0$, 
this system is described by two sets of second class constraints:
\ba\label{2.8}
\Omega^a_1&=& \pi^a_0, \nonumber \\
\Omega^a_2&=&(D^i \pi_i)^a + m^2(\eta) B^{0a},
\ea
where the covariant derivative is given by $(D^i \pi_i)^a 
= \partial^i \pi^a_i + \epsilon^{abc} B^{ib} \pi^c_i$.
Then, the constraints fully form the second class algebra as follows
\be\label{2.9}
\{\Omega^a_i(x),\Omega^b_j(y)\}=\Delta^{ab}_{ij}(x,y)\ee
with
\be\label{2.10}
\Delta^{ab}_{ij}(x,y)=\left(\begin{array}{cc}
                            0&-m^2(\eta)\delta^{ab}\\
          m^2(\eta)\delta^{ab}& \epsilon^{abc}\left(D^k \pi_k \right)^c
\end{array}\right)\delta^3(x-y).\ee

We now convert the second class system defined by the ``commutation  
relations'' (\ref{2.9}) to a first-class system 
at the expense of introducing  
additional degrees of freedom. Following refs. \cite{BF,BT}, 
we introduce  
auxiliary fields $\Phi^{1a}$ and $\Phi^{2a}$ corresponding to 
$\Omega^a_1$ 
and  $\Omega^a_2$, with the Poisson bracket
\be\label{2.11}
\left\{\Phi^{ia}(x),\Phi^{jb}(y)\right\}=\omega^{ij}_{ab}(x,y),
\ee
where we are free \cite{BF} to make the choice
\be\label{2.12}
\omega^{ij}_{ab}(x,y)=\epsilon^{ij}\delta_{ab}\delta^3(x-y).\ee
The first class constraints $\tilde\Omega^a_i$ are now constructed 
as a power  
series in the auxiliary fields,
\be\label{2.13}
\tilde\Omega^a_i=\Omega^a_i+\sum^\infty_{n=1}\Omega^{(n)a}_i,\ee
where $\Omega^{(n)a}_i(n=1,...,\infty)$ are homogeneous polynomials in the  
auxiliary fields $\{\Phi^{jb}\}$ of degree $n$, to be determined by the  
requirement that the constraints $\tilde\Omega^a_i$ be strongly involutive:
\be\label{2.14}
\left\{\tilde\Omega^a_i(x),\tilde\Omega^b_j(y)\right\}=0.\ee
Making the ansatz
\be\label{2.15}
\Omega^{(1)a}_i(x)=\int d^3y X^{ab}_{ij}(x,y)\Phi^{jb}(y)\ee
and substituting (\ref{2.15}) into (\ref{2.14}) leads to the condition
\be\label{2.16}
\int d^3 z d^3 z'X^{ac}_{ik}(x,z)\omega^{kl}_{cd}
(z,z')X^{bd}_{j\ell}(z',y)=-\Delta^{ab}_{ij}(x,y).\ee
With  the choice (\ref{2.12}) for $\omega^{ij}_{ab}(x,y)$, 
Eq. (\ref{2.16}) has  
(up to a natural arbitrariness) the solution
\be\label{2.17}
X^{ab}_{ij}(x,y)=\left(\begin{array}{cc}
m^2(\eta)\delta^{ab}&0\\
-{1\over 2}\epsilon^{abc}({\cal D}^k\pi_k)^c&\delta^{ab}\end{array}\right)
\delta^3(x-y).\ee

From the symplectic structure of Eq. (\ref{2.11}) with 
the choice (\ref{2.12})
for $\omega^{ij}_{ab}(x,y)$, we may identify the auxiliary 
fields with canonically
conjugated pairs. We make this explicit by adopting the notation 
\ba
(\Phi^{1a}, \Phi^{2a}) \Rightarrow (\theta^a, \pi^a_\theta). \nonumber
\ea

Substituting (\ref{2.17}) into (\ref{2.15}) as well as (\ref{2.13}), and  
iterating this procedure one finds the strongly involutive first class  
constraints to be given by
\ba
\tilde\Omega^a_1&=&\pi^a_0+m^2(\eta)\theta^a, \nonumber \\
\tilde\Omega^a_2&=& m^2(\eta)B^{0a}+
V^{ab}(\theta)({\cal D}^i\pi_i)^b +\pi_\theta^a.\label{2.18}
\ea
Here
\be\label{2.19}
\left({\it ad}\, \theta \right)^{ab}=\epsilon^{acb}\theta^c,
\ee
and
\be\label{2.20}
V(\theta)=\sum^\infty_{n=0}{{(-1)^n}\over{(n+1)!}}
({\it ad}\, \theta)^n,\ee
where ${\it ad} \,\theta$ denotes the Lie algebra valued field $\theta$ in 
the adjoint representation,  
${\it ad}\, \theta=\theta^aT^a$, with $T^c_{ab}=\epsilon^{acb}$. 
This completes the construction of the first class constraints.

\section{First class fields and Hamiltonian}
\setcounter{equation}{0}
Let ${\cal J}$ denote collectively the variables 
$(B^{\mu a},\pi^a_\mu,\eta,\pi)$ of the original phase space.
The construction of the first class Hamiltonian $\tilde H$ can be done
along similar lines as in the case of the constraints, by representing it as  
a power series in the auxiliary fields and requiring
$\{\tilde\Omega^a_i,\tilde H\}=0$ subject to the condition 
$\tilde H[{\cal J},\theta^a = \pi^a_\theta = 0]=H_C$.
We shall  follow here  a somewhat different
path \cite{KK,BBN,PP,KR} by noting that any 
functional of first class fields $\tilde  
{\cal J}$ corresponding to ${\cal J}$ in the extended phase space
will also be first class.
This leads us to the identification $\tilde H=H_C[\tilde {\cal J}]$. The  
``physical'' fields $\tilde{\cal J}$ are obtained 
as a power series in the  
auxiliary fields $(\theta^a, \pi^a_\theta)$ by requiring them to be 
strongly involutive:  
$\{\tilde\Omega^a_i,\tilde{\cal J} \}=0$. 
The iterative solution of these  
equations involves the use of (\ref{2.12}) and (\ref{2.17}) 
and leads to an  
infinite series which  can be compactly written in terms of 
${\it ad}\, \theta$  defined in (\ref{2.19}) as
\ba\label{3.1}
\tilde B^{0a}&=& B^{0a}+\frac{1}{m^2(\eta)}\pi_\theta^a-
\frac{1}{m^2(\eta)}\left(U^{ab}(\theta)-V^{ab}(\theta)\right)
({\cal D}^i\pi_i)^b,\nonumber\\
\tilde B^{ia}&=& U^{ab}(\theta) B^{ib}+V^{ab}(\theta)\partial^i\theta^b,
\nonumber \\
\tilde\pi^a_0&=&\pi^a_0+m^2(\eta)\theta^a, \nonumber\\
\tilde\pi_i^a &=& U^{ab}(\theta)\pi_i^b, \nonumber\\
\tilde\eta &=& \eta, \nonumber\\
\tilde\pi &=& \pi + 2m(\eta)B^{0a}\theta^a,
\ea
where $V(\theta)$ has been defined in (\ref{2.20}) 
and $U(\theta)$ is given by
\be\label{3.2}
U(\theta)=\sum^\infty_{n=0}{{(-1)^n}\over{n!}}({\it ad} \,\theta)^n
=e^{-{\it ad}\,\theta}.\ee
We now observe that the first class constraints (\ref{2.18})
can also be written in terms of the physical fields as
\ba\label{3.3}
\tilde\Omega^a_1&=&\tilde\pi^a_0,\nonumber\\
\tilde\Omega^a_2&=&(\widetilde{{\cal D}^{i} \pi_i})^a
                   + m^2(\eta)\tilde B^{0a}. 
\ea
Comparing with the second class constraints $\Omega^a_i$
in Eq. (\ref{2.8}), we see that the first class constraints  
(\ref{3.3}) are just the second class constraints written in terms of the  
physical variables. Correspondingly, we take the first class Hamiltonian  
density $\tilde{\cal H}_C$ to be given by the second class one (\ref{2.7}),  
expressed in terms of the physical fields:
\be\label{3.4}
\tilde{\cal H}_C= {1\over 2} (\tilde\pi^a_i)^2 
            + {1\over 2} (\tilde\pi)^2
            + {1 \over 4} (\tilde G^a_{ij})^2 
            + {1\over 2} m^2(\tilde\eta) (\tilde B^{ia})^2  
            + {1\over 2} m^2(\tilde\eta) (\tilde B^{0a})^2  
            + {1\over 2} (\partial_i \tilde\eta)^2
            - V(\tilde\eta) 
            - \tilde B^{0a}\tilde \Omega_2^a.
\ee
It is important to notice that any Hamiltonian weakly equivalent to  
(\ref{3.4}) describes the same physics since the observables of 
the first class formulation must be first class themselves. 
Hence we are free to add to  
$\tilde{\cal H}$ any terms proportional to the first class constraints.

For later comparison, we explicitly rewrite the above Hamiltonian by making
use of the expressions in Eq. (\ref{3.1}): 
\ba\label{3.5}
\tilde{\cal H}_C &=& {1\over 2} (\pi^a_i)^2 
                + {1\over 2} \left(\pi + 2m(\eta)B^{0a}\theta^a \right)^2
                + {1 \over 4} (G^a_{ij})^2 
                + {1 \over 2} m^2(\eta) 
  \left(U^{ab}(\theta)B^{ib}+V^{ab}(\theta)\partial^i\theta^b\right)^2 
        \nonumber \\
                &+& {1 \over 2} (\partial_i \eta)^2
                - V(\eta) 
                + {1 \over 2m^2(\eta)} \left(({\cal D}^i\pi_i)^a\right)^2 
                - {1 \over 2m^2(\eta)} (\tilde \Omega_2^a)^2.
\ea

\section{Lagrangian interpretation of the results}
\setcounter{equation}{0}

Let us first define the group valued field
\be\label{4.1}
g(\theta)=e^\theta,\quad \theta=\theta^at^a.\ee
Then, we have for a Lie algebra valued field $A=A^at^a$,
\ba\label{4.2}
-tr(t^ag^{-1}(\theta)Ag(\theta))=U^{ab}(\theta)A^b,\nonumber\\
-tr(t^ag^{-1}(\theta)\partial_\mu g(\theta))
=V^{ab}(\theta)\partial_\mu\theta^b,\ea
where the r.h.s. resumes in compact form an infinite series given by Eqs.
(\ref{2.20}) and (\ref{3.2}).
However, since the time component fields $\tilde B^{0a}$ in Eq. (\ref{3.1})
or Eq. (\ref{3.3}) can be at most rewritten as 
\be\label{4.3}
\tilde B^{0a}={1 \over m^2(\eta)} \left(\tilde \Omega^a_2 
              - U^{ab}(\theta)({\cal D}^i\pi_i)^b\right),
\ee
it is still difficult to directly understand the strong relation
of the gauge transform. On the other hand,
the spatial component fields $\tilde B^{ia}$ have a simple interpretation. 
From (\ref{4.2}) we see that the field $\tilde B^{ia}$ in the expression 
(\ref{3.1}) can be written in compact form as 
\be\label{4.4}
\tilde B^i = g^{-1} B^i g+g^{-1}\partial^ig,
\ee
which shows that $\tilde B^i$ is just the gauge transform of  $B^i$.
This suggests that as in the case of the non-abelian self-dual model 
\cite{KR}, our BFT results should be equivalent to the ones obtained 
by gauging $B^\mu$ in the Lagrangian (\ref{2.1}).
We now show that this is indeed the case.

Gauging the Lagrangian (\ref{2.1}) by making the substitution 
\be\label{4.5}
B^\mu \rightarrow \hat{B}^\mu = g^{-1} B^\mu g+g^{-1}\partial^\mu g,
\ee
we obtain
\be\label{4.6}
\hat{\cal L}(B,\eta,g)={\cal L}(B,\eta)+{\cal L}_{WZ},\ee
where
\be\label{4.7}
{\cal L}_{WZ}(B,\eta,g)=- m^2(\eta) \left( tr(B^\mu\partial_\mu gg^{-1})
+\frac{1}{2}tr(g^{-1}\partial^\mu g)^2 \right)
\ee
plays the role of the Wess-Zumino-Witten (WZW) term in
the gauge-invariant formulation of two-dimensional chiral gauge
theories \cite{BSV, HT, AAR}, and ${\cal L}(B,\eta)$ is the Lagrangian
of the second class system.

We then have for the momentum $\Pi$ conjugate to $g$,
\be\label{4.8}
\Pi^T=-m^2(\eta)(g^{-1}B^0+g^{-1}\partial^0gg^{-1}),\ee
where ``{\it T\,}'' denotes ``transpose''. 
The other canonical momenta
$\pi_\mu$ and $\pi$ are the same as before. Hence the primary constraints
are still of the form in Eq. (\ref{2.8}), $T_1^a=\pi_0^a$,
though the dynamics is a different one. The canonical Hamiltonian
corresponding to (\ref{3.5}) then reads, on the constraint surface
$\pi^a_0=0$,
\ba\label{4.9}
\hat {\cal H}_C&=& 
        {1 \over 2} (\pi^a_i)^2 + {1 \over 2} \pi^2       
         -\frac{1}{4}tr(G_{ij})^2 
        +{1 \over 2} m^2(\eta)(B^{ia})^2   \nonumber \\
&& 
      +m^2(\eta)tr(B^i\partial_ig g^{-1})
      -{1 \over 2} m^2(\eta)tr(g^{-1}\partial_i g)^2  
      +{1 \over 2} (\partial_i \eta)^2 - V(\eta)
\nonumber \\
&&   -\frac{1}{2m^2(\eta)}tr(\Pi^Tg)^2
     + tr B^0 ({\cal D}^i \pi_i - g\Pi^T ).  \ea
From (\ref{4.9}) we see that persistency in time of the primary constraints
$\pi^a_0=0$ implies secondary constraints associated with the Lagrange 
multipliers $B^{0a}$. We thus have two sets of first class constraints:
\ba\label{4.10}
T^a_1 &=& \pi^a_0 \approx 0, \nonumber \\
T^a_2 &=& ({\cal D}^i \pi_i)^a +tr(t^ag\Pi^T) \approx 0.
\ea

In order to establish the connection with the BFT results in section 3,
we must expand the terms involving $g$ and $\Pi^T$ into an infinite series
in the field $\theta$ using the identities (\ref{4.2}) and the following
properties of the Lie algebra and group valued functions, 
$V(\theta)$ and  $U(\theta)$ defined in Eqs. (\ref{2.20}) and (\ref{3.2}), 
respectively:
\ba\label{4.11}
V^{ab}(-\theta) &=& V^{ba}(\theta), 
          ~~U^{ab}(-\theta) \,=\, U^{ba}(\theta), \nonumber \\
          U^{ca}(\theta)V^{cb}(\theta)&=&V^{ab}(-\theta), 
          ~~U^{ac}(\theta)V^{bc}(\theta)\,=\,V^{ab}(\theta),\nonumber \\
&&U^{ca}(\theta)U^{cb}(\theta)=
                  U^{ac}(\theta)U^{bc}(\theta)=\delta^{ab}.
\ea

The WZ term (\ref{4.7}) can then be rewritten in form 
\be\label{4.12}
{\cal L}_{WZ} = m^2(\eta) B^{\mu a} U^{ca}(\theta) V^{cb}(\theta)
                \partial_\mu \theta^b
               + {1 \over 2} m^2(\eta) V^{ca}(\theta) V^{cb}(\theta)
                 \partial_\mu \theta^a \partial^\mu \theta^b.
\ee
From here we obtain for the momenta canonically conjugate to $\theta^a$
\be\label{4.13}
\pi'^{a}_{\theta} = m^2(\eta) \hat{B}^{0c} V^{ca}(\theta),
\ee
where  
\be\label{4.14}
\hat{B}^{0c} = U^{cb}(\theta) B^{0b} 
                 + V^{cb}(\theta)\partial^0 \theta^b  
\ee
is readily identified with the zero component of the gauged 
vector potential 
(4.5), and where we used the ``prime'' 
on $\pi^a_\theta$ in order to distinguish this momentum from the 
auxiliary field $\pi^a_\theta$ in Eq. (\ref{2.18}) 
introduced in the BFT construction in section 2.

It remains to establish the relation with the results
of section 3. Multiplying (\ref{4.8}) from the left with
$g$ and using (\ref{4.2}) we have
\ba\label{4.15}
tr(t^ag\Pi^T)&=&m^2(\eta)
\left(B^{0a}+V^{ab}(-\theta)\partial^0\theta^b\right)\nonumber\\
&=&m^2(\eta)\left(B^{0a}+U^{ca}
       (\theta)V^{cb}(\theta)\partial^0\theta^b\right).\ea
Making use of (\ref{4.14}), (\ref{4.11}) and (\ref{4.10}),
we obtain from (\ref{4.15}),
\be\label{4.16}
\hat B^{0a} = {1 \over m^2(\eta)}U^{ab}(\theta)tr(t^bg\Pi^T)
           ={1 \over m^2(\eta)}U^{ab}(\theta)(T_2 - 
                         {\cal D}^i\pi_i)^b.
\ee
Comparing this with (\ref{4.3}), we conclude 
$\hat B^{0a} \approx \tilde B^{0a}$ since 
$\hat B^{0a}$ and $\tilde B^{0a}$ are identical up to
additive terms proportional to the first class constraints. 
This establishes the weak equivalence of $\hat B^{\mu a}$
and $\tilde B^{\mu a}$.
Furthermore, 
combining (\ref{4.16}) and (\ref{4.13}) we have
\be\label{4.17}
\pi'^{a}_\theta=U^{ac}(\theta)V^{bc}(\theta)\,tr(t^bg\Pi^T).\ee
Using (\ref{4.10}) we rewrite this as
\be\label{4.18}
V^{ab}(\theta)T^b_2 =  \pi'^{a}_\theta + V^{ab}(\theta)
({\cal D}^i\pi_i)^b.
\ee
Performing the canonical transformation
\ba\label{4.19}
\pi^a_0 &\rightarrow& \pi^a_0 + m^2(\eta)\theta^a, \nonumber \\
\pi'^a_\theta &\rightarrow& \pi'^a_\theta + m^2(\eta)B^{0a}, 
   \nonumber \\
\pi &\rightarrow& \pi + 2m(\eta)B^{0a}\theta^a,
\ea
we see that the first class constraints 
$T^a_1 = \pi^a_0 \approx 0$ and 
$V^{ab}(\theta)T^b_2 \approx 0$ map into the constraints (\ref{2.18})
in the BFT construction.

It now remains to check the relation between $\hat {\cal H}_C$
and $\tilde {\cal H}_C$ as given by (\ref{4.9}) and (\ref{3.5}),
respectively. Making use of (\ref{4.14})
and (\ref{4.19}), expression
(\ref{4.9}) for $\hat {\cal H}_C$ may be rewritten in the 
following form in order to compare with $\tilde {\cal H}_C$ 
in Eq. (\ref{3.5})
\ba\label{4.20}
\hat{\cal H}_C &=& {1\over 2} (\pi^a_i)^2 
                + {1\over 2}  \left(\pi + 2m(\eta)B^{0a}\theta^a\right)^2
                + {1 \over 4} (G^a_{ij})^2 
                + {1 \over 2} m^2(\eta) 
   \left(U^{ab}(\theta)B^{ib}+V^{ab}(\theta)\partial^i\theta^b\right)^2 
\nonumber \\
                &+& {1 \over 2} (\partial_i \eta)^2
                - V(\eta)  
         + {1 \over 2m^2(\eta)} \left(({\cal D}^i\pi_i)^a\right)^2 
                + {1 \over 2m^2(\eta)} (T_2^a)^2 \nonumber \\
                &-& {1 \over m^2(\eta)}\left(m^2(\eta)B^{0a}
                    +  ({\cal D}^i\pi_i)^a\right) T^a_2. 
\ea
Then, we immediately obtain the equivalence relation 
$\hat {\cal H}_C \approx \tilde {\cal H}_C$ 
since $\hat {\cal H}_C$ is
identical with $\tilde {\cal H}_C$ up to additive terms
proportional to the first class constraints. 
We have thus arrived at a simple
interpretation of the results obtained in section 3.

Let us compare the Hamiltonian (\ref{3.5}) with the one given in 
Eq. (2.29) of \cite{BRR1}. Making use of (\ref{2.18}) we may rewrite 
$\tilde {\cal H}_C$ in the form
\be\label{4.21}
\tilde {\cal H}_C = {\cal H}_C + \Delta {\cal H}
\ee
where ${\cal H}_C$ has been defined in (\ref{2.7}) and $\Delta {\cal H}$ is
given by
\ba\label{4.22}
\Delta {\cal H}_C &=& 2 m(\eta) B^{0a}\theta^a 
                    \left(\pi + m(\eta)B^{0b}\theta^b \right) 
                    + m^2(\eta)V^{ab}(\theta)\partial^i\theta^b
                    \left(U^{ac}(\theta)B^{ic} + {1 \over 2}
             V^{ac}(\theta)\partial^i\theta^c \right) \nonumber \\
                    &+& {1 \over {2m^2(\eta)}}
                      \left(\tilde \Omega^a_2 - \pi^a_\theta
                             + (\delta^{ab} - V^{ab}(\theta))
                               ({\cal D}^i \pi_i)^b \right)
                    - {1 \over {2m^2(\eta)}} (\tilde \Omega^a_2)^2.
\ea
Taking the abelian limit, and comparing (\ref{4.21}) with the Hamiltonian
(2.29) of \cite{BRR1} one sees that the two Hamiltonians are weakly equal.  

The quantization of the first class system now proceeds following 
a standard
procedure \cite{FV} involving the construction of the BRST Hamiltonian 
$H_{BRST}$
as well as the unitarizing Hamiltonian $H_U$ obtained from $H_{BRST}$ 
by the
addition of a BRST exact term containing the gauge fixing (fermionic) 
function.
The steps are analogous to those taken in the abelian case of \cite{BRR1}, 
the only difference being that in the present case  
the algebra of the first class 
 constraints and Hamiltonian 
is strongly involutive, so that the BRST Hamiltonian coincides 
with the first class
Hamiltonian $\tilde {\cal H}_C$ in (3.5). In this respect the situation here 
differs from that of the abelian case discussed in \cite{BRR1}, 
where the first class
Hamiltonian $H'$ is in weak involution with the constraint 
$\tilde\Omega_1$.
However, the resulting difference in the unitarizing Hamiltonian 
can be shown not to affect
the corresponding path integral once part of the integration 
over the ghosts has been
performed. The recovery of the Higgs model in the unitary gauge 
for the choice $\chi_1^a = B^{0a}, \chi_2^a=\theta^a$ of the gauge 
fixing functions,
as well as its usual gauge
invariant formulation in terms of a complex scalar doublet 
$\phi = g(\theta)\frac{1}{\sqrt{2}}
(0,\eta+v)$ (with the BFT fields $\theta^a$ playing the role of 
the Goldstone
boson) in the Faddeev-Popov type gauges $\chi_1^a=B^{0a}$ and 
$\chi_2^a=f(B^i)$,
thus proceeds along the lines of the discussion of \cite{BRR1}.

\section{Conclusion}

The main objective of this paper was to provide a nontrivial,
 physically interesting  example
for the Hamiltonian embedding of a second-class theory
into a first-class one, following the systematic constructive
procedure of Batalin, Fradkin, and Tyutin \cite{BF,BT}.
Unlike the case of the abelian models discussed in the literature,
the first class Hamiltonian and secondary constraints generated
by this procedure are found to be given by an infinite power series
in the auxiliary fields in the extended phase space.
By explicitly summing this series we established the weak
equivalence with the corresponding quantities as obtained
by gauging the second class Lagrangian defining our 
SU(2) Higgs model in the unitary gauge, with
the auxiliary fields $\theta^a$ playing the role of the corresponding
gauge degrees of freedom. We thereby showed
that on the space of gauge-invariant functionals the Lagrangian
approach of refs. \cite{BSV,HT} for embedding second class
theories into a gauge theory is equivalent to the Hamiltonian
BFT approach. To do this we used the most economical
way for obtaining the desired results, by working
with group rather than Lie algebra valued fields in the
gauged Lagrangian.

\section*{Acknowledgment}
Two of the authors (Y.--W. Kim and Y.--J. Park) would
like to thank the Institut f\"ur Theoretische Physik for their
warm hospitality. The present study was partly supported by  
the Korea Research Foundation for (1996) overseas fellowship 
and the DFG--KOSEF Exchange Program, which made this 
collaboration possible.

\end{document}